\documentclass[10pt,a4paper]{article}
\usepackage[utf8]{inputenc}
\usepackage[english]{babel}
\usepackage{amssymb,graphicx}
\usepackage{amsmath,amsfonts}
\usepackage{amsthm}
\usepackage{framed}
\usepackage{fullpage}
\usepackage[makeroom]{cancel}
\usepackage[export]{adjustbox} 
\usepackage{feynmf}
\usepackage{microtype}
\usepackage{hyperref}

\newcommand*{\Nwarrow}{\rotatebox[origin=c]{14}{\(\xrightarrow{\hspace*{3.5cm}}\)}}
\newcommand{\beq}{\begin{equation}}
\newcommand{\eeq}{\end{equation}}
\newcommand{\beqa}{\begin{eqnarray}}
\newcommand{\eeqa}{\end{eqnarray}}
\newcommand{\CR}{\nonumber \\}

\newcommand{\m}{\mu}
\newcommand{\lam}{\lambda}

\def\be{\begin{eqnarray}}
\def\ee{\end{eqnarray}}
\def\nn{\nonumber}

\begin{document}

\begin{center}
\begin{small}
\hfill FIAN/TD-03/21\\
\hfill IITP/TH-04/21\\
\hfill ITEP/TH-07/21\\
\hfill MIPT/TH-04/21\\
\end{small}
\end{center}


\begin{center}
\begin{Large}
\fontsize{17pt}{27pt}
	\textbf{Duality in elliptic Ruijsenaars system\\
\vspace{0.3cm} and elliptic symmetric functions}
	\end{Large}
	
\bigskip \bigskip
\begin{large}A. Mironov$^{a,b,c}$\footnote{mironov@lpi.ru; mironov@itep.ru},
A. Morozov$^{d,b,c}$\footnote{morozov@itep.ru},
Y. Zenkevich$^{e,f,b,g,d}$\footnote{yegor.zenkevich@gmail.com}
 \end{large}
\\
\bigskip

\begin{small}
$^a$ {\it Lebedev Physics Institute, Moscow 119991, Russia}\\
$^b$ {\it ITEP, Moscow 117218, Russia}\\
$^c$ {\it Institute for Information Transmission Problems, Moscow 127994, Russia}\\
$^d$ {\it MIPT, Dolgoprudny, 141701, Russia}\\
$^e$ {\it SISSA, Triest, Italy}\\
$^f$ {\it INFN, Sezione di Trieste, IGAP, Trieste, Italy}\\
$^g$ {\it ITMP, Moscow, Russia}\\
\end{small}
 \end{center}
\medskip

\begin{abstract}
We demonstrate that the symmetric elliptic polynomials  $E_\lambda(x)$ originally discovered in the study of generalized Noumi-Shiraishi functions are eigenfunctions of the elliptic Ruijsenaars-Schneider (eRS) Hamiltonians that act on the mother function variable $y_i$ (substitute of the Young-diagram variable $\lambda$). This means they are eigenfunctions of the dual eRS system. At the same time, their orthogonal complements in the Schur scalar product, $P_R(x)$ are eigenfunctions of the elliptic reduction of the Koroteev-Shakirov (KS) Hamiltonians. This means that these latter are related to the dual eRS Hamiltonians by a somewhat mysterious orthogonality transformation, which is well defined only on the full space of time variables, while the coordinates $x_i$ appear only after the Miwa transform. This observation explains the difficulties with getting the apparently self-dual Hamiltonians from the double elliptic version of the KS Hamiltonians.
\end{abstract}

\section{Introduction}

The Calogero system \cite{Cal} and its various trigonometric \cite{CS,RS} and elliptic \cite{eC,RS} generalizations
is one of the basic examples in the theory of integrable systems,
which, after a discovery of integrable properties of QFT effective actions \cite{UFN2,GKMMM},
are at one of the focuses of modern theoretical studies.
The whole Calogero-Sutherland-Ruijsenaars-Schneider family is rich enough for study the $PQ$ dualities \cite{Rui,Fock,GM,M} (see also the spectral duality version in \cite{specdu}),
and one may hope this would be the first case where the self-dual
double-elliptic (dell) system will be clearly formulated and investigated \cite{BMMM,MM,ABMMZ,AMM}.
At the moment, we are still stack at the previous stage:
an explicit formulation of elliptic-trigonometric duality. Indeed,
though the $PQ$ duality has been explicitly formulated both at the classical \cite{Rui,Fock,GM,M} and quantum \cite{Et} levels for the rational and trigonometric systems, and, in principle, it is basically clear how it should be lifted to elliptic systems, any explicit formulas has been lacking so far in the multi-particle case. This paper is a big step towards a resolution of this problem.

We will discuss solely the quantum duality, where the task is to find a dual pair of functions (called mother functions) of two sets of variables, $x$ and $y$, and the two sets of Hamiltonians for which they are the eigenfunctions.
Within the Calogero-Sutherland-Ruijsenaars-Schneider family, one of the sets of variables is usually considered as the Miwa transform
of the ``time variables" $\{p_k\}$, while the other set corresponds to an analytical continuation from lengths of the Young diagram $\lambda$. The mother function functions themselves are then related to various symmetric polynomials $P_\lambda\{p\}$
from the Schur-Macdonald family.
The Hamiltonians are well known up to the eRS case, though their dual versions are not known.
However, using the $PQ$ duality, we are still able to find the eigenfunctions $E_\lambda\{p_k\}$ of these latter.
In this paper, we demonstrate that they are actually expressed through
the elliptic Macdonald functions constructed recently in \cite{AKMMdell3}.
A candidate for the dual Hamiltonian was proposed by P. Koroteev and Sh. Shakirov \cite{KS}\footnote{See also an earlier $N=2$ version in \cite{BH}.} (in fact, they proposed a dell version of the Hamiltonians, which is, however, not self-dual) but, as we explain in this paper, their eigenfunctions
are not exactly the same elliptic Macdonald polynomials $E_\lambda\{p_k\}$, but rather their orthogonal complements.
This sounds like an innocent difference, but in fact orthogonality is a somewhat
strange requirement within this context.
In particular, it is naturally defined in the space of time-variables $\{p_k\}$,
i.e. requires a lift from symmetric polynomials to a larger space
(and thus has no immediate analogue for the $y$ variables).
Even more important, this prevents the full KS Hamiltonians
from being explicitly self-dual so that, at least, one more step is necessary
to resolve the puzzle with the dell systems, and to clarify the apparent difference
between $H^{KS}$ and an explicit $N=2$ example from \cite{BMMM} as well as $N>2$ examples of \cite{MM,ABMMZ,AMM}.

It may seem straightforward to find a substitute of $H^{KS}$ that act on variables $x$
in explicitly known polynomials $E_\lambda (x)$ and leave them intact. However, one has to find as many such Hamiltonians as there are $x$ variables, and all these Hamiltonians have to commute. To put it differently, one has to construct a generating function of Hamiltonians that depends on an auxiliary spectral parameter $u$, and, for instance, in the case of KS Hamiltonians, this spectral parameter would be better to introduce in a tricky way \cite{GZ}: $H^{KS}$ are made in a non-local way from simpler auxiliary operators $O^{\rm trig}(u)$ depending already on the spectral parameter. As we demonstrate in this paper, the drop-out of $u$-dependence is a corollary of some $\theta$-function identities. In other words, the eigenvalue equations for $H^{KS}$ have an additional hidden structure, which is not explicit, and it is a question if and how it can be preserved by the change of the eigenfunctions and Hamiltonians. All this remains for the future work.

To summarize, in this paper, in section 2, we explain the notion of duality, and, in section 3, we define a bi-orthogonal pair of elliptic polynomial systems, $P_\lambda\{p_k\}$ and $E_\lambda\{p_k\}$. Further, we demonstrate that

\begin{itemize}
\item $E_\lambda\{p\}$ as functions of analytically continued $\lambda\longrightarrow \{y\}$
are the eigenfunctions of the eRS Hamiltonians (sec.\ref{elRu}).
\item $P_\lambda(x_i)$ as functions of $\{x_i\}$ are the eigenfunctions of
the $H^{KS}$ when the coordinate torus is degenerate (sec.\ref{triKS}).
\item The independence of the spectral parameters, which makes the  latter problem
self-consistent, follows from a family of theta-function identities (sec.\ref{triKS}).
\item Similar identities also explain the equivalence to solutions found previously in \cite{AKMMdell1} (sec.\ref{triKS}).
\end{itemize}
 All these relations are summarized in the diagram of sec.\ref{sec6}.
 \begin{itemize}
\item $E_\lambda\{p\}$ furnish a ``vertical'' representation of
  elliptic Ding-Iohara-Miki (eDIM) algebra, while the conjugate of
  trigonometric degeneration of $H^{KS}$ corresponds to a commuting
  subalgebra inside the eDIM algebra (sec.~\ref{sec:trig-ks-hamilt}).
\end{itemize}
We end with a short conclusion.

\paragraph{Notation.}
The Pochhammer symbols are defined to be
\be
(z;q)_\infty:=\prod_{n=0}^\infty (1-zq^n),\ \ \ \ \ \ \ (z;q,w)_\infty:=\prod_{n,m=0}^\infty (1-zq^nw^m)
\ee
We need both the odd $\theta$-function that we define as\footnote{It differs by a factor from the standard odd $\theta$-function \cite{BE}:
\be
\theta_1(u;\tau)={iw^{1/8}\cdot (w;w)_\infty\over \sqrt{z}}\cdot\theta_{w}(z)\Big|_{w=e^{2\pi i\tau},z=e^{2\pi i u}}
\ee}
\be
\theta_\omega(z):=(z;q)_\infty(q/z;q)_\infty
\ee
and one of the even $\theta$-functions
\be
\theta^{(e)}_\omega(z):=\sum_{k\in\mathbb{Z}}z^k\omega^{k^2}
\ee
The elliptic $\Gamma$-function is defined to be
\be
\Gamma(z;q,w):={(qw/z;w,q)_\infty\over (z;w,q)_\infty}=\exp\left[\sum_m{z^m-(wq/z)^m\over (1-q^m)(1-w^m)m}\right]
\ee
The elliptic Pochhammer symbol is defined as
\beq\label{Theta}
\Theta(z;q,\omega)_n: ={\Gamma(q^nz;q,\omega)\over\Gamma(z;q,\omega)}
\eeq
We also use the notation
\begin{equation}
  \label{eq:2}
  \psi(x) = \frac{\theta_{\omega} \left( \frac{q}{t} x \right) \theta_{\omega}(t x
    )}{\theta_{\omega}(x) \theta_{\omega}(q x)},\ \ \ \ \ \ \ \ \ \ \zeta(z) =   \frac{\theta_{\omega}(q^2z)\theta_{\omega}(tz)}{\theta_{\omega}(qtz)\xi(qz)}
\end{equation}
In particular, $\psi(x/q) = \psi(x^{-1})$ and $\psi(tz)=\zeta(z)\Big|_{q\leftrightarrow t}$.

Throughout the paper, we denote symmetric functions of variables $x_i$ as $M_\lambda(x_i)$, $P_\lambda(x_i)$, $E_\lambda(x_i)$, while these polynomials as functions of power sums $p_k:=\sum_ix_i^k$ are denoted as $M_\lambda\{p_k\}$, $P_\lambda\{p_k\}$, $E_\lambda\{p_k\}$. The elementary symmetric polynomials \cite{Mac} are denoted through $e_k$, and monomial symmetric polynomials \cite{Mac}, through $m_\lambda$.

\section{Duality}

\paragraph{The notion of quantum duality.}
The idea of $PQ$-duality of integrable many-body systems was first proposed by S. Ruijsenaars \cite{Rui} and was later discussed in \cite{Fock,GM,M} at the classical level and in \cite{Et} at the quantum level. While the classical $PQ$-duality is realized just in terms of Hamiltonians and their canonical transformations \cite{M}, the quantum duality requires the eigenvalue problem, i.e. the Hamiltonians are accompanied by the eigenfunctions from the very beginning. That is, if the eigenvalue problem for the Hamiltonian $\hat H_x$, which is an operator acting on the variable $x$, looks like
\be
\hat H_x\cdot\Psi(x;\lambda)={\cal E}(\lambda)\Psi(x;\lambda)
\ee
then the dual Hamiltonian acts on the variable $\lambda$:
\be
\hat H_\lambda^{(D)}\cdot\Psi^{(D)}(\lambda;x)={\cal E}^{(D)}(x)\Psi^{(D)}(\lambda;x)
\ee
so that $\Psi^{(D)}(\lambda;x)=\Psi(x;\lambda)$.
Here ${\cal E}$ and ${\cal E}^{(D)}$ are some fixed functions of the variables $x$ and $\lambda$ accordingly.
In the case of many-body integrable system there are several coordinates $x_i$, $i=1,...,N$
and the corresponding $\lambda_i$ are associated with the separated variables.
Integrability implies that, in this case, there are $N$ commuting Hamiltonians
and $N$ dual Hamiltonians.
In this context, one naturally considers the eigenfunction $\Psi_\lambda(x)$ as a function of the two continuous variables $x$ and $\lambda$. In the case of Hamiltonians from the Calogero-Sutherland-Ruijsenaars-Schneider family, the most informative are the Hamiltonians of the Dell system, which are elliptic both in the coordinates and in momenta, and are self-dual, i.e. $\hat H_k=\hat H_k^{(D)}$.

\paragraph{Duality in the trigonometric Ruijsenaars system.}
The simplest example is provided by the trigonometric Ruijsenaars system, which is self-dual. Its Hamiltonians are explicitly given by
\begin{equation}
  \label{eq:17}
      H_k^{(q,t)} = t^{\frac{k(k-1)}{2}}\sum_{
      \begin{smallmatrix}
        I \subset
        \{1,\ldots,N \}\\
        |I|=k
      \end{smallmatrix}
} \prod_{i \in I}\left[ \prod_{j \in \{1,\ldots,N\} \backslash I}  \frac{\left(1-
      \frac{t y_i}{y_j} \right)}{\left(1- \frac{y_i}{y_j}
    \right)} \right] \prod_{i \in I} q^{y_i \partial_{y_i}}.
\end{equation}
The eigenfunctions of these Ruijsenaars Hamiltonians are the Macdonald polynomials:
\be
H_k^{(q,t)}M_\lambda(x_i)={\cal E}^{(k)}_\lambda(q,t)\cdot M_\lambda(x_i)
\ee
where the eigenvalues are given by
\be
{\cal E}^{(k)}_\lambda(q,t)=e_k (q^{\lambda_i}t^{N-i})
\ee

Since this system is self-dual, i.e. the eigenfunction coincides with its dual, one just needs the property $\Psi(\lambda;x)=\Psi(x;\lambda)$. This is guaranteed by the duality relation of the Macdonald polynomials:
\begin{equation}
{M_\mu(q^{\lambda_i}t^{-i})\over M_\mu(t^{-i})}={M_\lambda(q^{\mu_i}t^{-i})\over M_\lambda(t^{-i})}
\end{equation}
In fact, the normalization coefficient $M_\lambda(t^{-i})$ can be chosen in a different form, which we will need in further elliptic generalization.

Consider the set of $x_i:=t^{-i}$, $i=1,...,\ldots, N$ and the Young diagram such that $l_\lambda\le N$. Then,
\beq
p_k={1\over t^{kN}}{1-t^{kN}\over 1-t^k}
\eeq
and
$$
M_\lambda(x_i)=t^{\nu_\lambda} \prod_{i<j\le N}{(q^{\lambda_i-\lambda_j}t^{j-i};q)_\infty (t^{j-i+1};q)_\infty\over
(q^{\lambda_i-\lambda_j}t^{j-i+1};q)_\infty (t^{j-i};q)_\infty}=
$$
\beq\label{Norm}
=t^{\nu_\lambda} \prod_{i<j\le N}{(q^{\lambda_i-\lambda_j}t^{j-i};q)_\infty \over
(q^{\lambda_i-\lambda_j}t^{j-i+1};q)_\infty }\cdot\prod_{i=1}^N{(t^i;q)_\infty\over (t;q)_\infty}\sim \hbox{qDim}_\lambda(A=q^N)
\eeq
where qDim$_\lambda$ is the Macdonald dimension, and $\nu_\lambda:=\sum_i(i-1)\lambda_i$. In particular, one may put $N=\infty$ (assuming that $|t|<1$).

\section{Elliptic Macdonald polynomials}

In this section, we define the two conjugate systems of symmetric polynomials that are the main players in the paper. We call them elliptic Macdonald polynomials. They come as a particular case of the generalized Noumi-Shiraishi (GNS) polynomials \cite{AKMMdell3} defined as a certain
truncation of an explicit series expression. These polynomials form a basis in the space of symmetric functions. An essential property of both these systems of polynomials is that their coefficients are expressed not though separate $\theta$-functions, but through their peculiar combinations $\zeta(z)$ and $\psi(z)$ differing by the permutation of $q$ and $t$.

\subsection{$P_\lambda\{p\}$ polynomials}

The basic systems of polynomials, $P_\lambda^{(q,t,\omega)}(x_i)$ was defined in \cite{FOS,AKMMdell2,AKMMdell3} (in the latter reference, this system is obtained as a particular case of the GNS polynomials when $\xi(z)=\theta_\omega(z)$), and can be described in the following way generalizing the Noumi-Shiraishi representation
of the Macdonald polynomials \cite{NS}:
\be\label{GNS}
\boxed{
P_\lambda^{(q,t,\omega)}(x_i)=\prod_{i=1}^N x_i^{\lambda_i}
\cdot\sum_{m_{ij}}{\cal C}^\lambda_n(m_{ij}|q,t)\prod_{1\le i<j\le N}\left({x_j\over x_i}\right)^{m_{ij}}}
\ee
where $m_{ij}=0$ for $i\ge j$, $m_{ij}\in \mathbb{Z}_{\ge 0}$, the number of lines in the Young diagram $\lambda$ does not exceed $N$, and
\be
\begin{array}{lc}
{\cal C}^\lambda_n(m_{ij},|q,t):=&
\prod_{k=2}^n\prod_{1\le i<j\le k}{
\displaystyle{\Theta\Big(q^{\lambda_j-\lambda_i+\sum_{a>k}(m_{ia}-m_{ja})}t^{i-j+1};q,\omega\Big)_{m_{ik}}}
\over \displaystyle{\Theta\Big(q^{\lambda_j-\lambda_i+\sum_{a>k}(m_{ia}-m_{ja})}qt^{i-j};q,\omega\Big)_{m_{ik}}}}\times\nn\\
&\times\prod_{k=2}^n\prod_{1\le i\le j<k}{\displaystyle{
\Theta\Big(q^{\lambda_j-\lambda_i-m_{jk}+\sum_{a>k}(m_{ia}-m_{ja})}qt^{i-j-1};q,\omega\Big)_{m_{ik}}}
\over \displaystyle{\Theta\Big(q^{\lambda_j-\lambda_i-m_{jk}+\sum_{a>k}(m_{ia}-m_{ja})}t^{i-j};q,\omega\Big)_{m_{ik}}}}
\end{array}
\label{c}
\ee
As it was explained in \cite[sec.3]{AKMMdell2}, {\bf $P_\lambda^{(q,t,\omega)}(x_i)$ are symmetric polynomials}, which is a consequence of a series of non-trivial $\theta$-function relations. They form a ring with the properties described in \cite{AKMMdell3}. The first few of these polynomials are
\begin{align}
P_{[1]}^{(q,t,\omega)}\{p_k\} &= p_1 \nn \\
P_{[1,1]}^{(q,t,\omega)}\{p_k\} &= \frac{p_1^2-p_2}{2}
\nn\\
P_{[2]}^{(q,t,\omega)}\{p_k\} &= \frac{2-\zeta(1)}{2}p_2 + {\zeta(1)\over 2}p_1^2
\nn \\
P_{[1,1,1]}^{(q,t,\omega)}\{p_k\} &= \frac{p_3}{3} - \frac{p_2p_1}{2} + \frac{p_1^3}{6}\nn\\
P_{[2,1]}^{(q,t,\omega)}\{p_k\} &= p_2p_1+ {\zeta(1)+\zeta(t)-3\over 3}p_3-{\zeta(1)+\zeta(t)-2\over 2}p_2p_1
+{\zeta(1)+\zeta(t)\over 6}p_1^3
\nn \\
P_{[3]}^{(q,t,\omega)}\{p_k\} &= \left(1-\zeta(q)\zeta(1) + \frac{\zeta(q)\zeta(1)^2}{3}\right)p_3
+ \zeta(q)\zeta(1)\left(1-\frac{\zeta(1)}{2}\right)p_2p_1 + \frac{\zeta(q)\zeta(1)^2}{6}p_1^3
\end{align}
More examples can be found in \cite{AKMMdell3}.

\subsection{$E_\lambda\{p\}$ polynomials\label{secE}}

\paragraph{Orthogonality and $E$ polynomials.} Let us define a conjugate system of polynomials in the following way. Denote $\chi_{\lambda\Delta}$ the coefficients of the $p$-expansion of the
\be\label{ex1}
P_\lambda^{(q,t,\omega)}\{p_k\}=\sum_\Delta \chi_{\lambda\Delta}(q,t,\omega)\cdot p_\Delta
\ee
Then, the set of polynomials
\be\label{ex2}
P_\lambda^{(q,t,\omega)\perp}\{p_k\}=\sum_\Delta \chi^{-1}_{\Delta \lambda}(q,t,\omega)\cdot{p_\Delta\over z_\Delta}
\ee
with $\chi^{-1}$ being the inverse matrix, $p_\Delta:=\prod_{i=1}\Delta_i$ and $z_\Delta$ being the standard symmetric factor of the Young diagram (order of the automorphism) \cite{Fulton}, is orthogonal,
\be\label{or}
\Big<P_\lambda^{(q,t,\omega)}\Big|P_\mu^{(q,t,\omega)\perp}\Big>=\delta_{\lambda\mu}
\ee
w.r.t. to the measure
\be\label{orm}
\left<p_\Delta\Big|p_{\Delta'}\right>\ =\ z_\Delta\delta_{\Delta,\Delta'}
\ee

Note that, in the Macdonald limit $\omega\to 0$,
\be
P_\lambda^{(q,t,\omega\to 0)\perp}\{p_k\}=P_{\lambda^\vee}^{(t,q,\omega\to 0)}\{(-1)^{k+1}p_k\}
\ee
This suggests to define a system of symmetric polynomials
\be
\boxed{
E_{\lambda}^{(q,t,\omega)}\{p_k\}:=P_{\lambda^\vee}^{(t,q,\omega)\perp}\{(-1)^{k+1}p_k\}
}
\ee
such that
\begin{equation}
  \label{eq:6}
  E_{\lambda}^{(q,t,0)}\{p_k\} = P_\lambda^{(q,t,0)}\{p_k\}=M_{\lambda}^{(q,t)}\{p_k\}
\end{equation}

The first few of $E_{\lambda}^{(q,t,\omega)}(p_n)$ polynomials are
given by
\begin{align}
  \label{eq:1}
  E_{[1]}^{(q,t,\omega)}\{p_k\} &= p_1\\
  E_{[1,1]}^{(q,t,\omega)}\{p_k\} &= \frac{1}{2} (p_1^2 - p_2)\\
  E_{[2]}^{(q,t,\omega)}\{p_k\} &= \frac{2-\psi(t)}{2} p_1^2 +
  \frac{\psi(t)}{2}p_2,\\
  E_{[1,1,1]}^{(q,t,\omega)}\{p_k\} &=\frac{p_3}{3} - \frac{p_2p_1}{2} + \frac{p_1^3}{6}\\
  E_{[2,1]}^{(q,t,\omega)}\{p_k\} &= \frac{1}{6} (3 - \psi(t) \psi(t^2))
  p_1^3 + \frac{1}{2} (\psi(t) \psi(t^2) - 1) p_1 p_2 -
  \frac{1}{3}\psi(t) \psi(t^2) p_3\\
  E_{[3]}^{(q,t,\omega)}\{p_k\} &= \left( 1 - \frac{\psi(t)}{2}  -
    \frac{\psi(q t)}{2}  + \frac{\psi(t)  \psi(t^2) \psi(q t)}{6}  \right)
  p_1^3 + \frac{\psi(t) +  \psi(q t) - \psi(t) \psi(t^2) \psi(qt)}{2} p_1 p_2 + \frac{\psi(t) \psi(qt) \psi(t^2)}{3} p_3
\end{align}

\paragraph{Ring structure.}
\label{sec:rign-structure}
The $E_{\lambda}^{(q,t,\omega)}\{p_k\}$ polynomials form a commutative ring (isomorphic to
the ring of symmetric polynomials):
\begin{equation}
  \label{eq:3}
  E_{\lambda}^{(q,t,\omega)} \{p_k\} E_{\mu}^{(q,t,\omega)}\{p_k\} = \sum_{\nu} N_{\lambda
    \mu}^{\nu}(q,t,\omega) E_{\nu}^{(q,t,\omega)}\{p_k\}.
\end{equation}
The coefficients $N_{\lambda \mu}^{\nu}(q,t,\omega)$ have some nice
properties:
\begin{enumerate}
\item {\bf They vanish whenever the corresponding Littlewood-Richardson
  coefficient vanishes.}
\item They are elliptizations of the $(q,t)$-Littlewood-Richardson coefficients for
  Macdonald polynomials, and, when the latter are factorized, they also factorize into products of theta functions.
\end{enumerate}

It is not hard to guess a formula for some classes of ring
coefficients. The important example is the Pieri rule:
\begin{equation}
  \label{eq:5}
  E_{[1]}^{(q,t,\omega)}\{p_k\} E_{\mu}^{(q,t,\omega)}\{p_k\} = p_1
  E_{\mu}^{(q,t,\omega)}\{p_k\} = \sum_{i=1}^{l(\mu)+1} \prod_{j = 1}^{i-1}
  \psi \left( \frac{q^{\mu_i}t^{1-i}}{q^{\mu_j}t^{1-j}} \right) E_{\mu +
  1_i}^{(q,t,\omega)}\{p_k\}.
\end{equation}
where $\mu + 1_i$ denotes a diagram obtained from the diagram $\mu$ by
adding one box in column $i$. If $\mu + 1_i$ is not a Young diagram,
the coefficient in front of $E_{\mu}^{(q,t,\omega)}\{p_k\}$ vanishes
automatically.

\section{$E_\lambda\{p\}$ as solutions to dual eRS system
\label{elRu}}

The eRS Hamiltonians are manifestly given by
\begin{equation}
  \label{eq:26}
    H_k^{(q,t,\omega)} = t^{\frac{k(k-1)}{2}}\sum_{
      \begin{smallmatrix}
        I \subset
        \{1,\ldots,N \}\\
        |I|=k
      \end{smallmatrix}
} \prod_{i \in I}\left[ \prod_{j \in \{1,\ldots,N\} \backslash I}  \frac{\theta_{\omega} \left(
      \frac{t y_i}{y_j} \right)}{\theta_{\omega} \left( \frac{y_i}{y_j}
    \right)} \right] \prod_{i \in I} q^{y_i \partial_{y_i}}.
\end{equation}
Their eigenfunctions $\Psi_\lambda(y_i)^{eRS}$ were conjectured in \cite[Eq.]{Shi} (see also (\ref{PsieRS}) below). In this section, we are going to construct the dual of these eigenfunctions following the pattern of sec.\ref{dua}. That is, we use the duality relation
\be\label{dua}
\Psi_\lambda^{eRS}(y_i)\Big|_{y_i=q^{\mu_i}t^{-i}}=\Psi_\mu^{(D)}(x_i)\Big|_{x_i=q^{\lambda_i}t^{-i}}
\ee
where the dual functions $\Psi^{(D)}_\lambda(x_i)$ are eigenfunctions of the dual eRS Hamiltonians (yet to be evaluated). We demonstrate below that
\be
\Psi_\mu^{(D)}(x_i)\sim E_\mu^{(q,t,\omega)}\{p_n\}
\ee
by checking that, upon a proper choice of the normalization factor, it satisfies the eigenvalue equations with the Hamiltonians (\ref{eq:26}) w.r.t. to the variables $y_i=q^{\mu_i}t^{-i}$. Here $p_n=\sum_i x_i^n$.

\paragraph{eRS eigenfunctions.}
\label{sec:ellipt-ruijs-schn}

The first eRS Hamiltonian is given by
\begin{equation}
  \label{eq:7}
  H_1^{(q,t,\omega)} = \sum_{i=1}^N \prod_{j \neq i} \frac{\theta_{\omega} \left(
      \frac{t y_i}{y_j} \right)}{\theta_{\omega} \left( \frac{y_i}{y_j}
    \right)} q^{y_i \partial_{y_i}},
\end{equation}
where $N$ is the number of particles.

The eRS Hamiltonian~\eqref{eq:7} is related to the Pieri's
rule~\eqref{eq:5}. One can see this by conjugating $H_1^{(q,t\omega)}$
with the function
\begin{equation}
  \label{eq:9}
F^{(q,t,\omega)}(\vec{y}) = \prod_{i<j} \left[y_j^{\beta} \Upsilon^{(q,t,\omega)} \left(
   \frac{y_i}{y_j} \right)\right]
\end{equation}
where $\beta = \frac{\ln t}{\ln q}$ and
\begin{equation}
  \label{eq:10}
  \Upsilon^{(q,t,\omega)}(y) = \prod_{n,m \geq 0}\frac{\left(1 - y \omega^n q^m \right) \left(1 - (ty)^{-1} \omega^{n+1} q^{m+1}
    \right)}{\left(1 - t y \omega^n q^m
    \right) \left(1 - y^{-1} \omega^{n+1} q^{m+1}
    \right)}=\prod_{i<j\le N}{\Gamma(yt;q)_\infty\over
\Gamma(y;q)_\infty }
\end{equation}
Up to a constant factor, it is an immediate elliptization of (\ref{Norm}).

The function $\Upsilon$ satisfies a simple difference equation
\begin{equation}
  \label{eq:24}
  \frac{\Upsilon^{(q,t,\omega)}(qy)}{\Upsilon^{(q,t,\omega)}(y)} = \frac{\theta_{\omega}(ty)}{\theta_{\omega}(y)},
\end{equation}
and thus
\begin{equation}
  \label{eq:25}
  F^{(q,t,\omega)}(\vec{y}) q^{y_i \partial_{y_i}}
  (F^{(q,t,\omega)}(\vec{y}))^{-1} = \prod_{j<i} \left[ t^{-1}
  \frac{\theta_{\omega}\left( \frac{t}{q} \frac{y_j}{y_i}
    \right)}{\theta_{\omega}\left( \frac{1}{q} \frac{y_j}{y_i}
    \right)}\right] \prod_{j>i} \left[\frac{\theta_{\omega}\left( \frac{y_i}{y_j}
    \right)}{\theta_{\omega}\left( t \frac{y_i}{y_j}
    \right)}\right] q^{y_i \partial_{y_i}}.
\end{equation}
From Eq.~\eqref{eq:25} we see that after conjugation half of the
factors in $\prod_{i \neq j}$ in $H_1^{(q,t,\omega)}$ cancel, while
the rest are ``doubled''. Finally, using the property of the Jacobi
theta function $\theta_{\omega}(y^{-1}) = -y^{-1} \theta_{\omega}(y)$
on the factors in the first square brackets in Eq.~\eqref{eq:25}, we
get the conjugated Hamiltonian, which is expressible through the
function $\psi(y)$ from Eq.~(\ref{eq:2}):
\begin{equation}
  \label{eq:11}
\tilde{H}_1^{(q,t\omega)} \stackrel{\mathrm{def}}{=} F^{(q,t,\omega)}(\vec{y}) H^{(q,t,\omega)}_1 (F^{(q,t,\omega)}(\vec{y}))^{-1}  = \sum_{i=1}^N \prod_{j<i} \frac{\theta_{\omega} \left(
      \frac{ty_i}{y_j} \right) \theta_{\omega} \left(
      \frac{qy_i}{t y_j} \right)}{\theta_{\omega} \left( \frac{y_i}{y_j}
    \right) \theta_{\omega} \left( \frac{q y_i}{y_j}
    \right)} q^{y_i \partial_{y_i}} = \sum_{i=1}^N \prod_{j<i} \psi \left( \frac{y_i}{y_j}
    \right) q^{y_i \partial_{y_i}} .
\end{equation}

If we send $N \to \infty$ and set the variables $y_i$ to discrete
values
\begin{equation}
  \label{eq:12}
  y_i = q^{\mu_i} t^{1-i}
\end{equation}
for some Young diagram $\mu$ we notice that the action of the
conjugated Hamiltonian~(\ref{eq:11}) coincides with the Pieri
rule~(\ref{eq:5}). Therefore, with proper normalizing constant,
elliptic Macdonald polynomials are eigenfunctions of the eRS
model. More specifically, the function $\Psi^{(q,t,\omega)}(p_n|y_i)$,

\bigskip

\hspace{-.6cm}\fbox{\parbox{17cm}{
\begin{equation}
  \label{eq:14}
 \Psi^{(q,t,\omega)}(p_k|y_i)|_{y_i = q^{\mu_i}t^{1-i}} =  (F^{(q,t,\omega)}(q^{\mu_i}t^{-i}))^{-1} E_{\mu}^{(q,t,\omega)}\{p_k\}
\end{equation}
satisfies
\begin{equation}
  \label{eq:15}
  H_1^{(q,t,\omega)} \Psi^{(q,t,\omega)}(p_k|y_i) = p_1 \Psi^{(q,t,\omega)}(p_k|y_i).
\end{equation}
}}

\bigskip

{\bf The variables $y_k$ are the mother-function arguments, which substitute the Young diagram index $\mu$ in
$E_\mu\{p\}$ and $E_\mu(x)$.}

\paragraph{Higher eRS Hamiltonians and more Pieri rules.}
\label{sec:high-ers-hamilt}
The eigenfunction $\Psi^{(q,t,\omega)}(p_n|y_i)$ is in fact an
eigenfunction of a whole infinite set of eRS Hamiltonians,
of which $H_1^{(q,t,\omega)}$ is only the first member.

Conjugating this Hamiltonian with the function
$F^{(q,t,\omega)}(\vec{y})$ we get
\begin{multline}
  \label{eq:27}
  \tilde{H}^{(q,t,\omega)}_n =F^{(q,t,\omega)}(\vec{y})
  H^{(q,t,\omega)}_k (F^{(q,t,\omega)}(\vec{y}))^{-1} =\\
  = \sum_{1 \leq i_1 <
    \ldots < i_k\leq N} \prod_{j_1<i_1} \psi \left( \frac{y_{i_1}}{y_{j_1}} \right) \prod_{
    \begin{smallmatrix}
      j_2<i_2\\
      j_2 \neq i_1
    \end{smallmatrix}
} \psi \left( \frac{y_{i_2}}{y_{j_2}} \right) \prod_{
    \begin{smallmatrix}
      j_3<i_3\\
      j_3 \neq i_1, i_2
    \end{smallmatrix}
} \psi \left( \frac{y_{i_3}}{y_{j_3}} \right)\cdots \prod_{
    \begin{smallmatrix}
      j_k<i_k\\
      j_k \neq i_1, \ldots, i_{k-1}
    \end{smallmatrix}
} \psi \left( \frac{y_{i_k}}{y_{j_k}} \right)
q^{y_1 \partial_{y_1} + \ldots + y_k \partial_{y_k}}.
\end{multline}

These Hamiltonians can also be understood as further Pieri
rules for elliptic Macdonald polynomials $E_{\lambda}^{(q,t,\omega)}$:
\begin{multline}
  \label{eq:28}
  E_{[1^n]}^{(q,t,\omega)}(p_k) E_{\mu}^{(q,t,\omega)}(p_k) =
  e_n(p_k) E_{\mu}^{(q,t,\omega)}\{p_k\} =\\
  =\sum_{i_1 <
    \ldots < i_k} \prod_{j_1<i_1} \psi \left( \frac{q^{\mu_{i_1}}t^{1-i_1}}{q^{\mu_{j_1}}t^{1-j_1}} \right) \prod_{
    \begin{smallmatrix}
      j_2<i_2\\
      j_2 \neq i_1
    \end{smallmatrix}
} \psi \left( \frac{q^{\mu_{i_2}}t^{1-i_2}}{q^{\mu_{j_2}}t^{1-j_2}}
\right) \cdots \prod_{
    \begin{smallmatrix}
      j_k<i_k\\
      j_k \neq i_1, \ldots, i_{k-1}
    \end{smallmatrix}
} \psi \left( \frac{q^{\mu_{i_k}}t^{1-i_k}}{q^{\mu_{j_k}}t^{1-j_k}}
\right) E_{\mu+1_{i_1}+\cdots+1_{i_k}}^{(q,t,\omega)}\{p_k\},
\end{multline}
where $e_k\{p_k\} = s_{[1^k]}\{p_k\}$ are elementary symmetric functions.
Correspondingly, the eigenvalues of the higher Hamiltonians are given
by elementary symmetric functions of $p_n$ variables
\begin{equation}
\boxed{
  \label{eq:29}
    H_k^{(q,t,\omega)} \Psi^{(q,t,\omega)}(p_k|y_i) = e_k\{p_k\} \Psi^{(q,t,\omega)}(p_k|y_i).
}
\end{equation}

\section{$P_\lambda(x)$ as the eigenfunctions of $H^{KS}$ in the elliptic-trigonometric limit
\label{triKS}}

The KS Hamiltonians are elliptic both in momenta and coordinates. Here we consider the case when the coordinate torus is degenerate so that the dependence on coordinates is trigonometric, we call this as ell-trig case. If one considers the trig-ell case instead, the KS Hamiltonians become the eRS Hamiltonians. Since the KS Hamiltonians are not self-dual, one should not expect that degenerating the coordinate torus would lead to the dual eRS system. Indeed, how we explain in this section, the corresponding eigenfunctions are the $P_\lambda(x)$ polynomials, which are conjugate to the wave functions of the dual eRS system, $E_\lambda^{(q,t,\omega)}$ w.r.t. to the Schur scalar product. As it follows from this scalar product, the conjugation can be realized with the substitutions $\displaystyle{p_k\to -{1\over k}{\partial\over\partial p_k}}$, which is not easy to realize on the subspace of finite number of the Miwa variables $x_i$.

\paragraph{$P_\lambda(x)$ as eigenfunctions of the ell-trig KS Hamiltonians.} One of the possibilities to proceed with the ell-trig KS Hamiltonians is to notice directly that their wave functions constructed in \cite[Eqs.(72)-(73)]{AKMMdell1} are nothing but the $P_\lambda(x)$ polynomials. In order to see this, one has to use rather tricky relations between the odd and even $\theta$-functions, the simplest of which is
\beq\label{49}
{\theta_w^{(e)}(q^2t^{-1})\theta_w^{(e)}(q^2tw^{-1})-t\theta_w^{(e)}(q^2t)\theta_w^{(e)}(q^2t^{-1}w^{-1})\over
\theta_w^{(e)}(q^2tw^{-1})\theta_w^{(e)}(t)-q\theta_w^{(e)}(tw^{-1})\theta_w^{(e)}(q^2t)}=
{\theta_w(q^2)\theta_w(t)\over\theta_w(qt)\theta_w(q)}
\eeq
The formulas in \cite{AKMMdell3} actually used brute force calculations. A smarter approach is to use the generating function of the KS Hamiltonians in the determinant form~\cite{GZ}, which can be written, in the ell-trig case, as
\begin{equation}
  \label{eq:8}
  \mathcal{O}^{\mathrm{trig}}(u) = \frac{1}{\Delta(x)}\det_{1\leq i,j
    \leq N} \left( x_i^{N-j}
  \theta_{\omega}(u t^{N-j} q^{x_i \partial_{x_i}}) \right).
\end{equation}
This matrix is triangular in the basis of monomial symmetric
polynomials $m_{\lambda}(\vec{x})$. The current
$\mathcal{O}^{\mathrm{trig}}(u)$ can be diagonalized with eigenvalues
\begin{equation}
  \label{eq:13}
  \kappa_{\lambda}(u) = \prod_{i=1}^N \theta_{\omega}(u q^{\lambda_i}t^{N-i}).
\end{equation}
However, since $\mathcal{O}^{\mathrm{trig}}(u)$ for different values
of $u$ \emph{do not commute,} the eigenfunctions in general
\emph{depend on $u$.} The exceptions here are the eigenfunctions
correspond to $\lambda = [1^k]$, $m_{[1^k]}(\vec{x}) =
s_{[1^k]}(\vec{x})$.

\begin{equation}
  \label{eq:19'}
\mathcal{O}^{\mathrm{trig}}(v) \left(
  \begin{array}{c}
    m_{[1,1]}\\
    m_{[2]}
  \end{array}
 \right)=
  \left(
    \begin{array}{cc}
       \theta_{\omega}(qtv)\theta_{\omega}(qv)  &0\\
       \theta_{\omega}(q^2tv)\theta_{\omega}(v) -  \theta_{\omega}(q^2v)\theta_{\omega}(tv) &  \theta_{\omega}(v)\theta_{\omega}(q^2tv)
    \end{array}
\right) \left(
  \begin{array}{c}
    m_{[1,1]}\\
    m_{[2]}
  \end{array}
 \right)
\end{equation}

To get commuting Hamiltonians one should take the ratio of the
generating functions $\mathcal{O}^{\mathrm{trig}}(u)$ at two
different values of $u$:
\begin{equation}
  \label{eq:18}
  H(v,u) = \mathcal{O}^{\mathrm{trig}}(v) (\mathcal{O}^{\mathrm{trig}}(u))^{-1}.
\end{equation}
Let us compute the first nontrivial eigenfunction of $H(v,u)$, which
should not depend on $u$ and $v$. The matrix of the operator $H(v,u)$
in the basis of $m_{\lambda}$ on the second level reads
\begin{equation}
  \label{eq:19}
H(v,u) \left(
  \begin{array}{c}
    m_{[1,1]}\\
    m_{[2]}
  \end{array}
 \right)=
  \left(
    \begin{array}{cc}
       \frac{\theta_{\omega}(qtv)\theta_{\omega}(qv)}{\theta_{\omega}(qtu)\theta_{\omega}(qu)} &0\\
      \frac{\theta_{\omega}(q^2u)\theta_{\omega}(tu)\theta_{\omega}(v)\theta_{\omega}(q^2tv)}
      {\theta_{\omega}(u)\theta_{\omega}(qu)\theta_{\omega}(qtu)\theta_{\omega}(q^2tu)} -
      \frac{\theta_{\omega}(q^2v)\theta_{\omega}(tv)}{\theta_{\omega}(qu)\theta_{\omega}(qtu)}&  \ \ \ \frac{\theta_{\omega}(v)\theta_{\omega}(q^2tv)}{\theta_{\omega}(u)\theta_{\omega}(q^2tu)}
    \end{array}
\right) \left(
  \begin{array}{c}
    m_{[1,1]}\\
    m_{[2]}
  \end{array}
 \right)
\end{equation}
We need {\it left} eigenfunctions, at the second level they are
\be
  \label{eq:20}
  \Psi_{[1,1]} &=& m_{[1,1]}\nn\\
  \Psi_{[2]} &=&m_{[2]} +
  \frac{\theta_{\omega}(q^2)\theta_{\omega}(t)}{\theta_{\omega}(q)\theta_{\omega}(qt)} \cdot m_{[1,1]}
  = m_{[2]} + \zeta(1)\cdot m_{[1,1]}
  \label{eq:22}
\ee
The eigenfunctions indeed are independent of $u$ and $v$. This depends
on the following identity:
\begin{equation}
  \label{eq:21}
  \frac{\theta_{\omega}(u)\theta_{\omega}(q^2tu) \theta_{\omega}(q^2v)
    \theta_{\omega}(tv) -
    \theta_{\omega}(q^2u)\theta_{\omega}(tu)\theta_{\omega}(v)\theta_{\omega}(q^2tv)}
    {\theta_{\omega}(u)\theta_{\omega}(q^2tu) \theta_{\omega}(qv)
    \theta_{\omega}(qtv) - \theta_{\omega}(qu)\theta_{\omega}(qtu)\theta_{\omega}(v)\theta_{\omega}(q^2tv)}
  = \frac{\theta_{\omega}(q^2)\theta_{\omega}(t)}{\theta_{\omega}(q)\theta_{\omega}(qt)} = \zeta(1)
\end{equation}
We notice that the eigenfunctions~\eqref{eq:20}, \eqref{eq:22} are
precisely the $P_\lambda\{p_k\}$ polynomials.

Likewise, at the next level the {\it left} eigenfunctions are
\be
  \Psi_{[1,1,1]} &=& m_{[1,1,1]}\nn\\
  \Psi_{[2,1]} &=&m_{[2,1]} +\Big(\zeta(1)+\zeta(t)\Big)\cdot m_{[1,1,1]}\nn
  \\
  \Psi_{[3]} &=&m_{[3]} + \zeta(1)\zeta(q)\cdot m_{[2,1]} + \zeta(1)^2\zeta(q) \cdot m_{[1,1,1]}
\ee
and
\be
  \Psi_{[1,1,1,1]} &= m_{[1,1,1,1]}
  \nn\\
  \Psi_{[2,1,1]} &=m_{[2,1,1]} +\Big(\zeta(1)+\zeta(t)+\zeta(t^2)\Big)\cdot m_{[1,1,1,1]}
  \nn\\
  \Psi_{[2,2]} &=m_{[2,2]} + \zeta(1)\Big(\zeta(1)+\zeta(t)\Big)\cdot m_{[2,1,1]} + \zeta(t) \cdot m_{[1,1,1,1]}
  \nn\\
  \!\!\!\!\!\! \Psi_{[3,1]} \ \ \ \ \ \ \ \  &  \!\!\!\!\!\!\!\!\!\!\!\!\!\!\!\!\!\!\!\!\!\!\!\!
  =m_{[3,1]} +\zeta(1)\cdot m_{[2,2]} +  \zeta(1)\Big(\zeta(q)+\zeta(qt)\Big)\cdot m_{[2,1,1]}
  +  \zeta(1)\Big(\zeta(1)\zeta(q)+\zeta(1)\zeta(qt)+\zeta(t)\zeta(qt)\Big) \cdot m_{[1,1,1,1]}
  \nn\\
  \!\!\!\!\!\!\!\!\Psi_{[4]}\ \ \ \ \ \ \ \ \ \ &\!\!\!\!\!\!\!\!\!\!\!\!\!\!\!\!\!\!\!\!\!\!\!\!
  =m_{[4]} + \zeta(1)\zeta(q)\zeta(q^2)\cdot m_{[3,1]} +\zeta(1)\zeta(q)^2\zeta(q^2)\cdot  m_{[2,2]} + \zeta(1)^2\zeta(q)^2\zeta(q^2)\cdot m_{[2,1,1]} + \zeta(1)^3\zeta(q)^2\zeta(q^2) \cdot m_{[1,1,1,1]}
\nn  \label{eq:22'}
\ee
which are exactly the $P_\lambda\{p_k\}$ polynomials. We conjecture that this statement is true at all levels so that
\begin{equation}
\boxed{
  \label{eq:23}
  \hat H(v,u) \, P_\lambda^{(q,t,\omega)}(\vec{x}) = \prod_{i=1}^N
  \frac{\theta_{\omega}(v q^{\lambda_i}t^{N-i})}{\theta_{\omega}(u
    q^{\lambda_i}t^{N-i})}\cdot P_\lambda^{(q,t,\omega)}(\vec{x})
    }
\end{equation}
Note that counterparts of (\ref{eq:21}) behind (\ref{eq:22'}) are more involved,
the simplest one being
$$
  \frac{\theta_{\omega}(t^mu)\theta_{\omega}(t^{m+1}q^{m+2}u) \theta_{\omega}(t^mq^{m+2}v)
    \theta_{\omega}(t^{m+1}v) -
    \theta_{\omega}(t^mq^{m+2}u)\theta_{\omega}(t^{m+1}u)\theta_{\omega}(t^mv)\theta_{\omega}(t^{m+1}q^{m+2}v)}
    {\theta_{\omega}(t^mu)\theta_{\omega}(t^{m+1}q^{m+2}u) \theta_{\omega}(t^mqv)
    \theta_{\omega}(t^{m+1}q^{m+1}v) - \theta_{\omega}(t^mqu)\theta_{\omega}(t^{m+1}q^{m+1}u)\theta_{\omega}(t^mv)\theta_{\omega}(t^{m+1}q^{m+2}v)}
    =
$$
\begin{equation}
  \label{eq:21'}
  =   \prod_{i=0}^m \zeta(q^{i})
\end{equation}
The reason is that an elliptic function with two given poles is fully defined by one of its zeroes
and the overall scale. This is also behind the identities like (\ref{49}). There are plenty of other relations  associated with multiple poles.

\section{ELS-functions \cite{AKMMdell2}, dualities and conjugation\label{sec6}}

In \cite{AKMMdell2}, there was introduced and discussed the ELS-function defined in the following way:
\beq
\mathfrak{P}_N { (x_i ; p \vert y_i ; s \vert q,t,\omega)}
:= \sum_{\vec\lam} \prod_{i,j=1}^N
\frac{\mathcal{N}_{\lam^{(i)}, \lam^{(j)}}^{(j-i)} (t y_j/y_i \vert q,s,\omega)}
{\mathcal{N}_{\lam^{(i)}, \lam^{(j)}}^{(j-i)} (y_j/y_i \vert q,s,\omega)}
\prod_{\beta=1}^N \prod_{\alpha \geq 1} \left( \frac{p x_{\alpha + \beta}}{t x_{\alpha + \beta -1}} \right)^{\lam_\alpha^{(\beta)}},
\label{EG}
\eeq
where
\beqa \label{EGfactor}
\mathcal{N}_{\lam, \mu}^{(k)} (u \vert q, s, \omega)
&:=&
\prod_{j \geq i \geq 1 \atop j - i \equiv k~(\mathrm{mod}~n)} \Theta(uq^{-\mu_i + \lam_{j+1}} s^{-i +j} ; q,\omega)_{\lam_j - \lam_{ j+1}} \CR
&&~~~\times
\prod_{j \geq i \geq 1 \atop j  - i \equiv -k-1~(\mathrm{mod}~n)}
 \Theta (uq^{\lam_i - \m_j} s^{i - j -1} ; q,\omega)_{\m_j - \m_{j+1}}.
\eeqa
and $p$ is another elliptic parameter.

This function is a lift of the $P_\lambda^{(q,t,\omega)}(\vec{x})$ polynomial with the Shiraishi functor \cite{AKMMdell3} and is conjectured to play an essential role in description of the double elliptic systems. Here we note that various trig-ell and ell-trig limits of the ELS-function are related by dualities. Indeed, the $P_\lambda^{(q,t,\omega)}(\vec{x})$ polynomial is obtained from $\mathfrak{P}_N { (p^{N-i}x_i ; p \vert s^{N-i}y_i ; s \vert q,t,\omega)}$ in the $p\to 0$ limit. Indeed, consider the limit
\be\label{Pf}
\mathfrak{F}_{N} (x_i \vert y_i \vert q, t, \omega):=
\lim_{p \to 0} \mathfrak{P}_N { (p^{N-i}x_i ; p \vert s^{N-i}y_i ; s \vert q,t,\omega)}
\ee
Then,
\be\label{Ppf}
P_\lambda^{(q,t,\omega)}(x_i)=\prod_{i=1}^Nx_i^{\lambda_i}\cdot
\mathfrak{F}_{N} (x_i \vert q^{\lambda_i}t^{N-i} \vert q, {q\over t}, \omega)
\ee
At the same time, this is related to the $\omega\to 0$ limit of the ELS-function, \cite{Shi}
\beq
\mathfrak{E}_N (x_i; p  \vert y_i; s \vert q,t):=\lim_{\omega\to 0}\mathfrak{P}_N { (x_i ; p \vert y_i ; s \vert q,t,\omega)}
\eeq
via the formula \cite{FOS}
\beq\label{FOSid}
\mathfrak{F}_{N} (y_i \vert x_i \vert q, t, p)=\mathfrak{N}\cdot
\mathfrak{E}_N (p^{-i/N} x_i; p^{1/N}  \vert t^{i/N} y_i; t^{-1/N} \vert q,t)
\eeq
with the normalization constant
\beq
\mathfrak{N} :=
\left( \frac{(p;p)_\infty (pt; q, p)_\infty}{(qp/t ; q, p)_\infty} \right)^N
\prod_{1 \leq i < j \leq N} \frac {\Gamma(q x_i/x_j; q, p)}{\Gamma(t x_i/x_j; q, p)}
\prod_{1 \leq i < j \leq N} \frac {(q x_i/x_j; q)_\infty}{(t x_i/x_j; q)_\infty}
\eeq
Moreover, this mother function $\mathfrak{E}_n (x_i; p  \vert y_i; s \vert q,t)$ allows one to construct also the eigenfunctions of the eRS Hamiltonians: as was conjectured in \cite{Shi}, the function
\begin{equation}\label{PsieRS}
\Psi_\mu^{eRS}(x_i;q,t,p):=\prod_{i=1}^Nx_i^{\mu_i}\cdot\lim_{s\to 1}
{\mathfrak{E}_N (p^{N-i\over N}x_i; p^{1/N}  \vert y_i; s   \vert q,{q\over t})\over \alpha( p   \vert y_i; s \vert q,{q\over t})}\Big|_{y_i=q^{\mu_i}t^{N-i}}
\end{equation}
with $ \alpha( p   \vert y_i; s \vert q,t)$ being the constant term in the Shiraishi function $\mathfrak{E}_n (x_i; p  \vert y_i; s \vert q,t)$, which does not depend on $x_i$ (this normalization constant is necessary, since otherwise the limit of $s\to 1$ is singular), is an eigenfunction of the eRS Hamiltonian:
\be
H_1^{(q,t,\omega)}\Psi_\mu^{eRS}(x_i;q,t,p)=\Lambda_\mu(q,t,p)\cdot \Psi_\mu^{eRS}(x_i;q,t,p)
\ee
More discussion of this equation can be found in \cite{LNSh}. 

These formulas imply the conjecture that the eigenfunctions of the full KS Hamiltonians can be obtained from the limit $s\to 1$ of the ELS-functions $\mathfrak{P}_N { (x_i ; p \vert y_i ; s \vert q,t,\omega)}$:
\begin{equation}\label{ELSKS}
\boxed{
\Psi_\mu^{KS}(x_i;q,t,p,\omega):=\prod_{i=1}^Nx_i^{\mu_i}\cdot\lim_{s\to 1}
{\mathfrak{P}_N (p^{N-i\over N}x_i; p^{1/N}  \vert y_i; s   \vert q,{q\over t},\omega)\over \alpha^f( p   \vert y_i; s \vert q,{q\over t},\omega)}\Big|_{y_i=q^{\mu_i}t^{N-i}}
}
\end{equation}
with some normalization constant $\alpha^f( p   \vert y_i; s \vert q,{q\over t},\omega)$ that makes the expression non-singular in the $s\to 1$ limit. This should be the case, since the ELS-function describes the Nekrasov function in the full $\Omega$-background, while $s=e^{-2\pi\epsilon_2}\to 1$ \cite[Eq.(95)]{AKMMdell2} describes its Nekrasov-Shatashvili limit, and this the Nekrasov-Shatashvili limit that describes the quantum integrable system. As usual, this limit is singular and requires some accurate normalization. We have checked this conjecture in the first terms of expansion in the elliptic parameters $p$ and $\omega$.

\bigskip

Hence, we finally come to the diagram

\bigskip

\hspace{-.6cm}\fbox{\parbox{17cm}{
$$
\begin{array}{cccccr}
&&\mathfrak{P}_N { (x_i ; p \vert y_i ; s \vert q,t,\omega)}&&&\\
\\
&^{p\to 0}{\swarrow}&&{\searrow}^{\omega\to 0}&\\
\\
\mathfrak{F}_{N} (x_i \vert y_i \vert q, t, \omega)&&\stackrel{(\ref{FOSid})}{\longleftrightarrow}&&
\mathfrak{E}_N (x_i; p  \vert y_i; s \vert q,t)&\\
\\
^{(\ref{Ppf})}\downarrow &&&&^{(\ref{PsieRS})}\downarrow\\
\\
P_\lambda^{(q,t,\omega)}(x_i)&\stackrel{\rm conj.}{\longleftrightarrow}&
E_\lambda^{(q,t,\omega)}(x_i)&\stackrel{\rm duality}{\longleftrightarrow}&\Psi_\mu^{eRS}(x_i;q,t,p)
&\phantom{aaaa}\fbox{\checkmark}\\
\\
^{\rm sec.5}\uparrow&&^{\rm sec.4}\uparrow&&^{\cite{Shi}}\uparrow&\\
\\
\hbox{ell-trig KS Hamiltonians}&&\hbox{dual eRS Hamiltonians}&&\hbox{eRS Hamiltonians}&\\
&&(\hbox{unknown})&&\\
\uparrow&&&&\uparrow\\
&&\Nwarrow&\nwarrow&\\
\hbox{full KS Hamiltonians}&&&&\hbox{self-dual dell Hamiltonians}\\
&&&&(\hbox{unknown})
\end{array}
$$
}}

\bigskip

The checked line was the point of our interest in this paper.  Its main drawback is the mysterious orthogonality relation between the first two eigenfunctions  and the lack of explicit relation between the last two apart from (\ref{dua}).  The top of the table (the ELS-functions) is conjectured to provide solutions to the full KS Hamiltonian eigenproblem via (\ref{ELSKS}), which implies the next, most interesting step to be done: since (\ref{ELSKS}) looks providing eigenfunctions of  the non-self-dual KS Hamiltonians, what are appropriate self-dual functions?

\section{Elliptic DIM algebra and ell-trig KS Hamiltonians}
\label{sec:trig-ks-hamilt}

\paragraph{New view on the vertical Fock  representation of the elliptic DIM algebra.}
\label{sec:repr-ellipt-dim}
Elliptic DIM algebra (eDIM)~\cite{Saito} is generated by currents
$x^{\pm}(z)$ and $\psi^{\pm}(z)$ satisfying commutation relations
with an elliptic structure function. To keep the presentation brief, we
do not give here these relations, which can be found e.g.\
in~\cite{Ghoneim:2020sqi}. It was also found in~\cite{Ghoneim:2020sqi}
that the eDIM algebra can be rewritten as a direct sum of
the trigonometric DIM algebra and an additional Heisenberg subalgebra. We
adopt this view in this exposition.

\begin{figure}[h]
  \centering
\includegraphics[width=7cm]{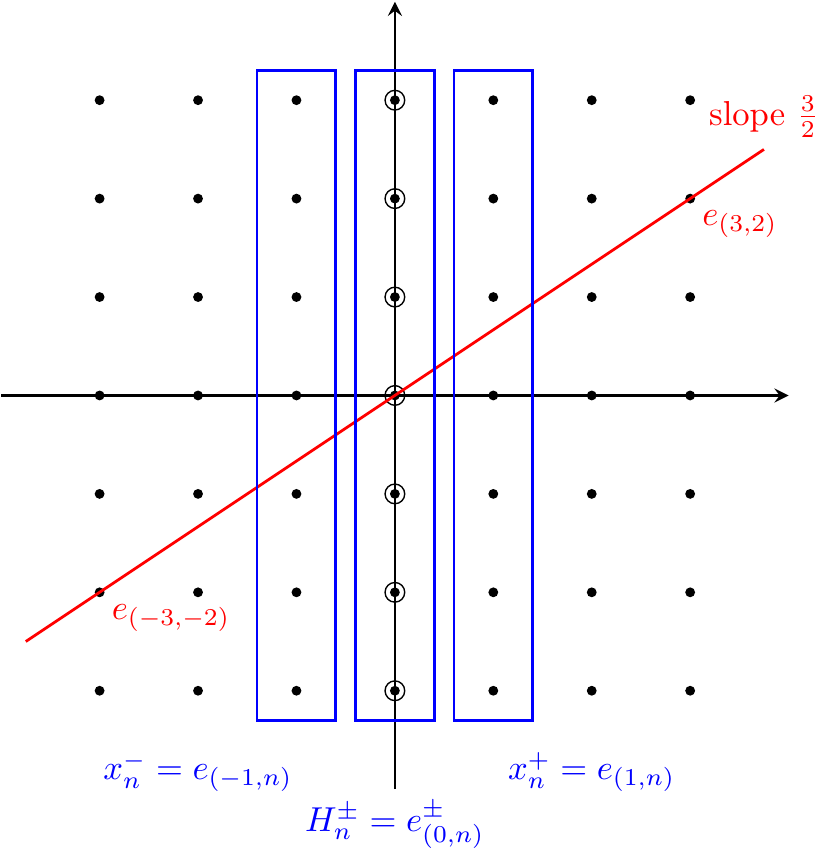}
\caption{The grading lattice of eDIM algebra. The generator at
  vertex $(n,m)$ is denoted by $e_{(n,m)}$. The generating currents
  are framed in blue, $x^{\pm}(z) = \sum_{n \in \mathbb{Z}} e_{(\pm 1,
    n)}z^{-n}$ and $\psi^{\pm}(z) = \exp \sum_{n\neq 0}H^{\pm}_n
  z^{-n} = \exp \sum_{n\neq 0}e^{\pm}_{(0,n)} z^{-n}$. Notice that the
  vertices on the vertical line unlike the others contain \emph{two}
  generators $e^{\pm}_{(0,n)}$. The red line denotes an example of a
  Heisenberg subalgebra of slope $\frac{3}{2}$.}
\label{fig:1}
\end{figure}

The eDIM algebra is bi-graded with generators lying in a
$\mathbb{Z}^2$ lattice, as shown in Fig.~\ref{fig:1}. Every node
contains a single generator\footnote{(Elliptic) DIM algebra is defined
  as a deformation of the universal enveloping algebra of the Lie algebra
  $qW_{1+\infty}$, hence it also contains nonlinear combinations of
  generators like $e_{(1,1)}^2$ sitting at the same node $(2,2)$ as
  $e_{(2,2)}$. When we say ``one generator per node'' we mean only
  ``primitive'' generators not expressible as products of the lower
  ones.}  $e_{(n,m)}$, except for the nodes on the vertical line which
contain two $e^{\pm}_{(0,m)}$, which is marked by additional small circles. 
The extra generators $e^{+}_{(0,m)}$ on
the vertical axis form a decoupled Heisenberg subalgebra commuting
with the rest of the DIM algebra.

There are two central charges $C_{1,2}$ in the algebra, $C_2$
associated with the ``elliptic'' direction (vertical in
Fig.~\ref{fig:1}), and $C_1$ associated with the ``trigonometric'' one
(horizontal in Fig.~\ref{fig:1}). The central charges control the
commutation relations in the Heisenberg subalgebras of the eDIM algebra which correspond to lines of rational slopes in the
$\mathbb{Z}^2$ lattice. The Heisenberg subalgebra corresponding to the
slope $\frac{a}{b}$ is spanned by the generators $e_{(na,nb)}$ (with an
additional $\pm$ index in case of the vertical subalgebra)
\begin{equation}
  \label{eq:32}
  [e_{(na,nb)}, e_{(ma,mb)}] \sim (C_1^{-nb} C_2^{na} - C_1^{nb} C_2^{-na}) \delta_{n+m,0}
\end{equation}
Therefore, if, for a particular representation, the ratio $\frac{\ln
  C_1}{\ln C_2}$ is rational, then there exists a \emph{commuting}
Heisenberg subalgebra inside the eDIM algebra of slope $\frac{\ln
  C_1}{\ln C_2}$.

We are particularly interested in the vertical Fock representation
$\mathcal{F}_{q,t^{-1}}^{(0,1)}(u)$ of the eDIM algebra with $C_1 = 1$, $C_2 = \sqrt{t/q}$. In this representation,
the \emph{both} vertical subalgebras spanned by $e_{(0,n)}^{\pm}$ are
commutative. The $e^{+}_{(0,n)}$ generators turn out to be irrelevant
in this representation: they commute both between themselves and with
the rest of the algebra. We set them to zero. The rest of the
generators can be expressed through the horizontal Heisenberg
subalgebra $e_{(n,0)}$, which acts freely on the Fock space. The
states of the representation are labelled by Young diagrams, e.g.\
$|\mu,u\rangle$.

In the vertical representation, the generating currents act as
follows~\cite{Zhu, Wang}
\begin{align}
  \label{eq:16}
  x^{+}(z) |\mu,u\rangle &= \frac{1}{1-q^{-1}}\sum_{i=1}^{l(\mu)+1}
  \delta \left( \frac{q^{\mu_i}t^{1-i} u}{z} \right) \prod_{j<i}\psi
  \left( \frac{z}{q^{\mu_j}t^{1-j} u} \right) |\mu+1_i,u\rangle,
  \\
  x^{-}(z) |\mu,u\rangle &= \frac{1}{1-q} \sqrt{\frac{q}{t}} \sum_{i=1}^{l(\mu)} \delta
  \left( \frac{q^{\mu_i-1}t^{1-i}u}{z} \right) \widetilde{\prod}_{j>i}\psi \left(
    \frac{q^{\mu_j-1}t^{1-j}u}{z} \right) |\mu+1_i,u\rangle,
  \\
  \psi^{+}(z) |\mu,u\rangle &= \sqrt{\frac{q}{t}} \widetilde{\prod}_{j=1}^{\infty} \psi
  \left(
    \frac{q^{\mu_j}t^{1-j}u}{z} \right) |\mu,u\rangle =
  \sqrt{\frac{q}{t}} \frac{\theta_{\omega}
    \left( \frac{t}{q}\frac{u}{z} \right)}{\theta_{\omega} \left( \frac{u}{z} \right)} \prod_{(i,j) \in \mu}
  g(q^{j-1} t^{1-i}u/z))|\mu,u\rangle, \label{eq:39}\\
  \psi^{-}(z) |\mu,u\rangle &= \sqrt{\frac{t}{q}} \widetilde{\prod}_{j=1}^{\infty} \psi
  \left(
    \frac{q z}{q^{\mu_j}t^{1-j}u} \right) |\mu,u\rangle =
\sqrt{\frac{t}{q}}  \frac{\theta_{\omega}
    \left( \frac{q}{t}\frac{z}{u} \right)}{\theta_{\omega} \left( \frac{z}{u} \right)} \prod_{(i,j) \in \mu}
  g(q^{1-j} t^{i-1}z/u))^{-1}|\mu,u\rangle.\label{eq:40}
\end{align}
where $g(x) = \frac{\theta_{\omega}(q^{-1} x) \theta_{\omega}(t x) \theta_{\omega}((q/t) x)}{\theta_{\omega}(q x) \theta_{\omega}(t^{-1} x) \theta_{\omega}((t/q) x)}$, and the tildes over the
infinite products indicate a regularization. One can immediately notice
that the action of the \emph{zero mode} $x^{+}_0$ of the current
$x^{+}(z)$ coincides with the Pieri rule~\eqref{eq:5} for the elliptic
symmetric functions $E_{\lambda}^{(q,t,\omega)}\{p_n\}$. Indeed, we
have
\begin{multline}
  \label{eq:36}
 \oint \frac{dz}{z} x^{+}(z) |\mu,u\rangle = \frac{1}{1-q^{-1}}\sum_{i=1}^{l(\mu)+1}
  \oint \frac{dz}{z} \delta \left( \frac{q^{\mu_i}t^{1-i} u}{z} \right) \prod_{j<i}\psi
  \left( \frac{q^{\mu_i}t^{1-i} u}{q^{\mu_j}t^{1-j} u} \right)
  |\mu+1_i,u\rangle =\\
  = \frac{1}{1-q^{-1}}\sum_{i=1}^{l(\mu)+1}
  \prod_{j<i}\psi
  \left( \frac{q^{\mu_i}t^{1-i} u}{q^{\mu_j}t^{1-j} u} \right)
  |\mu+1_i,u\rangle.
\end{multline}
This observation suggests that the \emph{horizontal} Heisenberg
subalgebra of the eDIM algebra spanned by $e_{(n,0)}$ acts on
the vertical Fock representation in the same way as the power sum
variables $p_n$ and $n \frac{\partial}{\partial p_n}$ act on
$E_{\lambda}^{(q,t,\omega)}\{ p_n \}$:
\begin{equation}
  \label{eq:37}
  e_{(n,0)}E_{\lambda}^{(q,t,\omega)} \{ p_n\} \sim
  \begin{cases}
    p_n E_{\lambda}^{(q,t,\omega)} \{ p_n\}, & n >0,\\
    |n| \frac{\partial}{\partial p_{|n|}}E_{\lambda}^{(q,t,\omega)} \{ p_n\}, & n<0.
  \end{cases}
\end{equation}

It is natural to ask: how does the \emph{vertical} Heisenberg
subalgebra act on $E_{\lambda}^{(q,t,\omega)}\{ p_n \}$? The answer is
given by the \emph{conjugates} of the ell-trig KS Hamiltonians. As we have shown in sec.~\ref{triKS}, the
conjugates of $E_{\lambda}^{(q,t,\omega)}\{ p_n \}$, called
$P_{\lambda}^{(q,t,\omega)}\{ p_n \}$, are eigenfunctions of the
these Hamiltonians, so $E_{\lambda}^{(q,t,\omega)}\{ p_n \}$
themselves are eigenfunctions of hypothetical conjugate ell-trig KS
Hamiltonians.

Let us note that the eigenvalues of the conjugate Hamiltonians
coincide. The conjugation we mean here is with respect to the standard
Schur scalar product~\eqref{orm}. Therefore, we can equivalently
consider the diagonal action of the KS Hamiltonians on
$P_{\lambda^{\vee}}^{(t,q,\omega)}\{ (-1)^{n+1} p_n\}$ instead of the
action of the conjugate KS Hamiltonians on
$E_{\lambda}^{(q,t,\omega)}\{ p_n\}$. To get the correct eigenvalues
$\psi^{\pm}(z)$, we cook up the following combination of
Hamiltonians~\eqref{eq:18}:
\begin{equation}
  \label{eq:41}
\chi^{\perp}_N(z) = \sqrt{\frac{q}{t}} H\left(t \frac{t^{1-N} u}{z},  \frac{t^{1-N}u}{z} \right) H\left(
    \frac{q}{t}  \frac{t^{1-N}u}{z}, q  \frac{t^{1-N}u}{z} \right) .
\end{equation}
The operator $\chi(z)$ is diagonal in the basis of polynomials
$P_{\lambda^{\vee}}^{(t,q,\omega)}(\vec{x})$. Expressing $\chi(z)$ in
terms of $p_n$, taking the (regularized) limit $N \to \infty$, and
then conjugating with respect to the standard scalar product, we get a
(hypothetical) operator $\chi(z)$, which is diagonal in the basis
$E_{\lambda}^{(q,t,\omega)}\{ p_n\}$, and whose eigenvalues are
\begin{equation}
  \label{eq:42}
  \chi(z) E_{\lambda}^{(q,t,\omega)}\{ p_n\} = \sqrt{\frac{q}{t}} \widetilde{\prod}_{j=1}^{\infty} \psi
  \left(
    \frac{q^{\mu_j}t^{1-j}u}{z} \right) E_{\lambda}^{(q,t,\omega)}\{ p_n\}.
\end{equation}
This is the operator representing the action of $\psi^{+}(z)$ of the DIM
algebra in the vertical Fock representation.

\paragraph{Ell-trig KS eigenvalues vs trigonometric RS system.}
It is curious that there exists a Hamiltonian with the same
eigenvalues as $\chi(z)$, but with eigenfunctions being ordinary
Macodnald polynomials. Indeed, consider the generating function of the
trigonometric RS Hamiltonians
\begin{equation}
  \label{eq:4}
  \Omega^{(q,t)}(u) = \sum_{k=0}^N (-u)^k H^{(q,t)}_k,
\end{equation}
where $H^{(q,t)}_k$ is given by (\ref{eq:17}), the trigonometric limit of
Eq.~\eqref{eq:26}.

The operators $\Omega^{(q,t)}(u)$ (unlike
$\mathcal{O}^{\mathrm{trig}}(u)$ from Eq.~\eqref{eq:8}) commute for
different $z$ and therefore can be diagonalized in a basis independent
of $u$. This basis is given by the Macdonald polynomials. The
eigenvalues $\kappa_{\lambda}^{(q,t)}(u)$ are
\begin{equation}
  \label{eq:30}
\kappa_{\lambda}^{(q,t)}(u)= \sum_{k=0}^N (-u)^k
  e_k(q^{\lambda_i} t^{N-i})  =
  \prod_{i=1}^N (1- u q^{\lambda_i} t^{N-i}),
\end{equation}
where $e_k = s_{[1^k]}$ denote the elementary symmetric functions.

We also notice that $M_{\lambda}^{(q,t)}(\vec{x})$ are eigenfunctions
of $H^{(q^{-1}t^{-1})}_k$:
\begin{equation}
  \label{eq:33}
  H^{(q^{-1}t^{-1})}_kM_{\lambda}^{(q,t)}(\vec{x}) = e_k(q^{-\lambda_i} t^{i-N})M_{\lambda}^{(q,t)}(\vec{x}).
\end{equation}

Having the commuting operators $\Omega^{(q,t)}(u)$ and
$\Omega^{(q^{-1},t^{-1})}(u)$ we can cook up an operator with
eigenvalues coinciding with that of $\mathcal{O}^{\mathrm{trig}}(u)$:
\begin{equation}
  \label{eq:31}
  \mathcal{O}'(u) = (\omega;\omega)_{\infty} \prod_{k\geq 0}
  \Omega^{(q,t)}(\omega^k u)
  \Omega^{(q^{-1},t^{-1})}\left( \frac{\omega^{k+1}}{u}\right).
\end{equation}
By construction the operator $\mathcal{O}'(u)$ is diagonalized by
Macdonald polynomials and the eigenfunctions are
\begin{equation}
  \label{eq:34}
  \mathcal{O}'(u) M_{\lambda}^{(q,t)}(\vec{x}) = \prod_{i=1}^N
  \theta_{\omega}(u q^{\lambda_i} t^{N-i}) M_{\lambda}^{(q,t)}(\vec{x}).
\end{equation}
Of course, we can take a ``ratio'' of two operators $\mathcal{O}'(u)$
for two different values of $u$ to get the eigenvalues as in
Eq.~\eqref{eq:18}:
\begin{equation}
  \label{eq:35}
  H'(v,u) = \mathcal{O}'(v) (\mathcal{O}'(u))^{-1}.
\end{equation}
Taking this idea one step further, we can write an operator with the
same eigenvalues as $\chi(z)$ from Eq.~\eqref{eq:42}:
\begin{equation}
  \label{eq:43}
  \chi'(z) = \sqrt{\frac{q}{t}} H'\left(t \frac{t^{1-N} u}{z},  \frac{t^{1-N}u}{z} \right) H'\left(
    \frac{q}{t}  \frac{t^{1-N}u}{z}, q  \frac{t^{1-N}u}{z} \right).
\end{equation}
Its eigenfunctions, however, are still ordinary Macdonald polynomials.

Thus, in this subsection we have obtained an operator $H'(v,u)$, whose
polynomial eigenfunctions are enumerated by Young diagrams with the
eigenvalues coinciding with those of the ell-trig KS
Hamiltonian. However, its eigenfunctions are not elliptic Macdonald
polynomials, but instead are given by the ordinary Macdonald
polynomials. This fits the observation
that the elliptic DIM algebra is in fact a nontrivial rewriting of the
ordinary trigonometric DIM algebra. It would be nice to find an
explicit transformation between the KS Hamiltonians and the
``untwisted'' Hamiltonians~\eqref{eq:35}. This transformation might
also explain why the elliptic Macdonald polynomials are not orthogonal
with respect to the Macdonald scalar product: the ``twist'' does not
respect the orthogonality, so, from the same set of orthogonal
Macdonald polynomials, one gets two sets of conjugate elliptic
Macdonald polynomials $E_{\lambda}^{(q,t,\omega)}\{p_k\}$ and
$P_{\lambda}^{(q,t,\omega)}\{p_k\}$.

\section{Conclusion}

As we see from the diagram of sec.\ref{sec6}, the choice of $E_{\lambda}^{(q,t,\omega)}\{p_k\}$ and $P_{\lambda}^{(q,t,\omega)}\{p_k\}$ for elliptic Macdonald polynomials looks rather elegant.
It is encouraging that $E_{\lambda}^{(q,t,\omega)}\{p_k\}$ are indeed distinguished as the eigenfunctions of the dual eRS Hamiltonians, while $P_{\lambda}^{(q,t,\omega)}\{p_k\}$ as the eigenfunctions of the ell-trig KS Hamiltonians.
The main open question which remains is what are the substitutes of the KS Hamiltonians
as functions of $p_k$: having these at hands, one could immediately make a conjugation $\displaystyle{p_k\to -{1\over k}{\partial\over\partial p_k}}$ in order to obtain dual eRS Hamiltonians, and, probably, further to construct similarly self-dual Hamiltonians.

The main point is that $H^{KS}$ have a nice formulation in terms of auxiliary operators $O^{\rm trig}(u)$,
which depend on an extra parameter $u$
and are not straightforwardly seen at the level of eigenfunctions themselves.
Drop-out of $u$-dependence follows from the theta-function identities,
which are not very convenient to deal with, which complicates a resolution of the problem.
Still, our results in this paper makes this problem very clear and transparent,
giving a hope that it will be finally resolved.
This will open a way to the search of explicitly self-dual formulation of the dell system.
A first task on this way is to clarify the conjecture of \cite{AKMMdell1} that the generalized Shiraishi function (the ELS-function \cite{AKMMdell2}) provides appropriate eigenfunctions:  either of $H^{KS}$ or of the truly self-dual Hamiltonians.
The $H^{KS}$ option extends the problem to search for a self-dual version of the Shiraishi construction.

\section*{Acknowledgements}

This work was supported by the Russian Science Foundation (Grant No.18-71-10073).

\end{document}